# Comments on frequency dependent AC conductivity in polymeric materials at low frequency regime

**Christos Tsonos**

Electronic Engineering Department, Technological Education Institute (TEI) of Sterea Ellada, 3rd Km O.N.R. Lamia-Athens, 35100 Lamia, Hellas, Greece

E-mail address: tsonos@teilam.gr

**Abstract**

The AC conductivity response in a broad frequency range of disordered materials is of great interest not only for technological applications, but also from a theoretical point of view. The Jonscher's power exponent value, and its temperature dependence, is a very important parameter in dielectric data analysis as well as the physical interpretation of conduction mechanisms in disordered materials. In some cases the power exponent of AC conductivity has been reported to be greater than 1 at the low frequency regime. This fact seems to contradict the universal dynamic response. The present work focuses on the analysis of dielectric spectroscopy measurements in polymeric materials, below ~100 MHz. The apparent power exponent $n$ gets values in the range $0<n<1$ and is directly related to the characteristics of mobile charges at shorter time scales, in the case of the occurrence of DC conduction and the slowest polarization mechanism that is due to the charge motions within sort length scales, in $\log\varepsilon''$-$\log\omega$ plot. The emergence of apparent $n$ values in the range $1\leq n\leq 2$, for a relatively narrow frequency range, may be attributed to an additional molecular dipolar relaxation contribution at higher frequencies, in $\log\varepsilon''$-$\log\omega$ plot. The appearance of apparent $n$ values in the range $1<n\leq 2$, can be assigned to the existence of a well defined minimum between DC conductivity contribution and a molecular dipolar dispersion or between two well separated dielectric loss mechanisms, in logε"-logω plots, above the crossover frequency. In these latter cases, the apparent power exponent *n* is merely related to the Havriliak-Negami equation shape parameters of the higher frequencies molecular dipolar relaxations.

# 1. Introduction

Ionic transport in disordered matter is a thermally activated process. It is believed that the diffusion of the mobile ions occurs via hopping between well-defined potential minima in the materials network. Information about these motions on different time and length scales can be obtained by using dielectric spectroscopy measurements and other techniques. In the classical dielectric region, bellow ~100 MHz, there are two main contributions in dielectric losses: conduction losses due to free charge carriers and dispersion losses due to dipolar relaxations or polarization mechanisms which are related to the short range charges motion. As it is well known, different disordered conductive materials such as polymers, semiconductors, ionic glasses and ceramics, exhibit a similar response to an applied electric field. Their behavior is described by the real part, $\sigma'(\omega)$, of the complex electrical conductivity $\sigma^*(\omega)$ which is usually called alternating current (AC) conductivity. Microscopically, at low frequencies, random diffusion of the charge carriers via activated hopping gives rise to a frequency independent conductivity. However, at higher frequencies, the AC conductivity, $\sigma'(\omega)$, exhibits dispersion increasing roughly in a power law fashion and eventually becoming almost linear at even higher frequencies. The AC conductivity can be expressed as:[1,2]

$$\sigma'(\omega) = \sigma_o + A\omega^n \qquad (1)$$

where $\omega$ is the angular frequency, $\sigma_o$ is the direct current (DC) conductivity, $A$ is a temperature dependent parameter and $n$ is a fractional exponent which take values $0<n<1$. Equation (1) is known as the universal dynamic response (UDR) or Jonscher's power law. Many theoretical approaches have been developed in order to explain the universal response described by Equation (1).[3-8] There have been also many independent theoretical attempts trying to conclude in a unique fractional exponent value close to 0.7 and justify the empirical universal Jonscher's power law.[9-12] However, experimentally obtained values for $n>1$ have been reported in a lot of cases concerning dielectric measurements for different types of disordered materials at a frequency range below 1 MHz. These materials include single crystals $K_3H(SeO_4)_2$,[13] mixed compounds $(NH_4)_3H(SO_4)_{1.42}(SeO_4)_{0.58}$,[14] chalcogenide semiconductor $Se_{0.9}In_{0.1}$,[15] $(Bi_{0.5}Na_{0.5})_{0.95}Ba_{0.05}TiO_3$ ceramic,[16] glassy $0.3(xLi_2O.(1-x)Li_2O)0.7B_2O_3$,[17] NiTPP thin films,[18] BPDA/ODA polyimide films,[19] $Pr_{0.6}Sr_{0.4}Mn_{0.6}Ti_{0.4}O_{3\pm\delta}$ perovskite,[20] silicon nitride MIS structures,[21] undoped $AgPO_3$ glass,[22] and lithium ferrite $Li_{0.5}MxFe_{2.5-x}O_4$ materials.[23]

Values of n>1 have been predicted from theoretical studies based on the assumption of an ideal network of conduction paths of various lengths.[24] Many experimental studies have reported their data to correspond to values of slope equal or higher than 1 at the high frequency regime.[25-29] Also, a nearly linear increase of $\sigma'(\omega_o)$ at higher frequencies corresponding to near constant losses (NCL) has been reported in the literature.[11,30] Dielectric measurements extending up to sufficiently high frequencies or low temperatures indeed reveal a superlinear behavior of $n$ exponent, as a result of vibrational modes contributions which take place at THz region.[25-27,29]

As mentioned above many models and theoretical approaches have been proposed for the explanation of AC conductivity response and the power exponent $n$, but their discussion and comparison goes beyond the scope of the present work. The high frequencies slope of AC conductivity is an important parameter in dielectric spectroscopy data analysis, while the values of $n$ exponent, as well as its temperature dependence, is of great importance for the physical interpretation of the various conduction mechanisms in disordered materials.[31-34] To the best of our knowledge there has not been published a study explaining where is due the appearance of *n*>1 values at the low frequency regime. This paper is based on the definitions of electromagnetic theory as well as the recorded dielectric response of polymeric materials as described by the widely used Havriliak and Negami (HN) equation. It focuses on the analysis of dielectric spectroscopy measurements. The purpose of the present study is to show the conditions under which the power law universal response of AC conductivity, exhibits sublinear and superlinear behavior at lower frequency regime, below ~100 MHz, in polymeric materials. This work demonstrates that UDR model can be misused because dipolar polarization processes can be active in parallel with charge carrier's mobility which is intended to be described by UDR. Taking into account the polarization processes and describing them by the HN function in the frequency domain it is shown that the apparent power low dependence of the real part of complex electrical conductivity, in a narrow frequency window, is actually determined by the low or high frequency tail of the HN mechanism. Although the present work has been based on fitting the data and performing calculations for polymeric materials, the analysis of this study could be extended to include other disordered systems whose dielectric loss mechanisms can be described by the HN equation.

## 2. Theory

The complex dielectric constant, $\varepsilon^*(\omega)$, is connected to the complex electrical conductivity, $\sigma^*(\omega)$, via the relation $\sigma^*(\omega) = j\omega\varepsilon_o\varepsilon^*(\omega)$. When the contribution of DC conductivity, $\sigma_o$, is subtracted from $\varepsilon^*(\omega)$ then $\sigma^*(\omega) = \sigma_o + j\omega\varepsilon_o\varepsilon_d^*(\omega)$, where $\varepsilon_d^*(\omega)$ represents the complex dielectric constant caused by dielectric loss mechanisms only. The real part of complex conductivity in the low frequency regime and in the absent of electrode polarization effects, is given by a similar to Equation (1) empirical equation which is $\sigma'(\omega) = \sigma_o \left(1 + (\omega/\omega_o)^n\right)$ where $n$ is the same fractional exponent. The characteristic frequency $\omega_o$ corresponds to the onset of the AC conductivity and is a characteristic hopping frequency of those ions contributing to the conductivity, while at frequency $\omega_o$ is $\sigma'(\omega_o) = 2\sigma_o$. Equation (1) and $\sigma'(\omega) = \sigma_o \left(1 + (\omega/\omega_o)^n\right)$ are equivalent when $\sigma_o = A\omega_o^n$, where the last relation gives the relationship between DC and AC conductivity.

The real part of the complex conductivity is given by the following relation:

$$\sigma'(\omega) = \sigma_o + \varepsilon_o \omega \varepsilon_d''(\omega) \quad (2)$$

The well known HN equation is widely used for the description of dielectric dispersions in polymeric materials:[35]

$$\varepsilon^*_d(\omega) = \varepsilon_\infty + \frac{\Delta\varepsilon}{(1+(j\omega\tau)^\alpha)^\beta} \quad (3)$$

where $\varepsilon_\infty$ is the dielectric constant at the high frequency limit and $\tau$ is a characteristic relaxation time related to a characteristic frequency $\omega_{HN}$ via the relation $\omega_{HN}\tau = 1$. The imaginary part is:

$$\varepsilon_d''(\omega) = \frac{\Delta\varepsilon \sin(\beta\varphi)}{\left[1 + 2(\omega/\omega_{HN})^\alpha \cos(\alpha\pi/2) + (\omega/\omega_{HN})^{2\alpha}\right]^{\beta/2}} \quad (4a)$$

where

$$\varphi = \arctan\left(\frac{(\omega/\omega_{HN})^\alpha \sin(\alpha\pi/2)}{1+(\omega/\omega_{HN})^\alpha \cos(\alpha\pi/2)}\right) \tag{4b}$$

$\alpha$, $\beta$ are shape parameters with values in the range $0 < \alpha, \beta \leq 1$. For the interpretation of HN shape parameters several models have been proposed.[36,37] The characteristic frequency $\omega_{HN}$ is connected to the loss peak frequency $\omega_{max}$ via the relation $\omega_{max} = A\omega_{HN}$ where:

$$A = \left(\frac{\sin(\alpha\pi/(2\beta+2))}{\sin(\alpha\beta\pi/(2\beta+2))}\right)^{1/\alpha} \tag{5}$$

Frequencies $\omega_{HN}$ and $\omega_{max}$ are equal in the cases of symmetrical behavior, $\alpha=\beta=1$ and $\beta=1$ (Debye and Cole-Cole). In general, both frequencies are close to each other depending on the shape parameters values. At the higher frequencies, $\omega >> \omega_{HN}$, Equation (4) becomes:

$$\varepsilon_d''(\omega) \cong \Delta\varepsilon\omega_{HN}^{\alpha\beta} \sin(\alpha\beta\pi/2)\omega^{-\alpha\beta} \tag{6}$$

while at lower frequencies, $\omega << \omega_{HN}$, Equation (4) becomes:

$$\varepsilon_d''(\omega) \cong \Delta\varepsilon\omega_{HN}^{-\alpha} \beta \sin(\alpha\pi/2)\omega^{\alpha} \tag{7}$$

## 3. Results and discussions

In the case of a disordered material with non-negligible conduction and without any dielectric loss mechanism contribution in the frequency spectra under study, both the AC conductivity and the real part of the complex dielectric constant, $\varepsilon'$, are frequency independent quantities according to Kramers-Kroning relations. The presence of a dielectric loss mechanism gives rise to dispersion in the AC conductivity, which increases roughly in a power law fashion. We consider now a polymeric material, which is characterized only by the contribution of DC conductivity, $\sigma_o$, and dielectric dispersion which presents a loss peak at frequency $\omega_{max}$ and dielectric strength $\Delta\varepsilon$.

When $\sigma_o \geq \varepsilon_o \omega \varepsilon''_d(\omega)$ at frequencies $\omega \leq \omega_{HN}$, then at $\omega > \omega_{HN}$ as $\omega$ increases the term $\varepsilon_o \omega \varepsilon''_d(\omega)$ at high frequency regime of $\varepsilon''_d(\omega)$ becomes larger than $\sigma_o$ and Equation (2) via Equation (6) is written as shown in a previously published work:[38]

$$\sigma'(\omega) \cong \sigma_o + \varepsilon_o \omega_{HN}^{\alpha\beta} \Delta\varepsilon \sin(\alpha\beta\pi/2) \omega^{1-\alpha\beta} \tag{8}$$

in whole frequency range. It is obvious that Equation (8) holds also when $\sigma_o \geq \varepsilon_o \omega \varepsilon''_d(\omega)$ up to frequencies higher than $\omega_{HN}$. The characteristic frequency $\omega_{HN}$ has been used here as a reference for reasons which will be made clear below. Equation (8) expresses what usually is observed and describes the majority of AC conductivity response in disordered materials. DC charge mobility, in disordered materials, always results to a polarization process, which is due to the charge motions within short length scales according the Random Barrier model.[9] This polarization mechanism obeys the Barton, Nakajima and Namikawa (BNN) relation and it is usually masked by DC conductivity effects or it appears as a shoulder. The power law dependence of the real part of complex electrical conductivity, as Equation (8) shows, characterized by an apparent exponent $n$ which is equal to $1-\alpha\beta$ and is directly related to the high frequency slope, *-αβ*, of the polarization process in $\log\varepsilon'' - \log\omega$ plots. Therefore, the faster components of the polarization mechanism form the value of high frequencies slope of AC conductivity. The HN dielectric function can be expressed as a superposition of individual Debye relaxations by introducing the distribution function of relaxation times. In this framework, a broader polarization process, which is equivalent to a broader distribution of its relaxation times, implies lower values of both HN shape parameters and a higher value of slope *1-αβ* (and vice versa).

Figure 1 represents the dielectric response of a polyurethane sample at 90 $^o$C. The obtained DC conductivity value is $\sigma_o$=3.6x10$^{-7}$ S/m, as extracted from the low frequency plateau of the AC conductivity versus frequency graph (Figure 1a). The imaginary part, $\varepsilon''$, of the complex dielectric constant as a function of frequency is also given in Figure 1b according to $\varepsilon''(\omega) = \sigma'(\omega)/\varepsilon_o\omega$. Figure 1b shows also the contribution of the polarization mechanism with HN fitting parameter values *α*=1, *β*=0.32, *Δε*=4.63 and $f_{HN}$= $\omega_{HN}$/2π=637 Hz, and DC conductivity term Kf$^{-1}$ with fitting parameter value K=6260,24. The dielectric loss mechanism is related to a polarization mechanism due to short range ions' motion and obeys the (BNN) relation.[38] At $\omega = \omega_{HN}$ the DC conductivity, $\sigma_o$, is about 10 times higher than the term $\varepsilon_o\omega\varepsilon''_d(\omega_{HN})$, and thus, Equation (8) can be used to describe the frequency dependence of AC conductivity in a very good approximation, as it can be seen in the simulation curve of Figure 1a. Generally, in polymeric materials with non negligible conduction, the mobile charges give rise for the existence of a polarization mechanism with maximum loss peak at frequencies lower than the onset characteristic frequency $\omega_o$, which implies that Equation (8) could be applied for the description of AC conductivity response in a very good approximation especially at low and high frequencies range.[38] This polarization mechanism is usually masked from the DC conductivity effect or appears as a shoulder in the $\log\varepsilon'' - \log\omega$ plots. It also satisfies the BNN relation which indicates that both AC and DC conductivity may arise from the same type of charge transport mechanism.[39,40] In these cases the apparent power exponent *n* is directly related to the characteristics of mobile ions at high frequencies or equivalent in shorter times scale. So, in the case where the dipolar-like polarization process originates in charge carriers accumulations within the disorder polymeric matrix, it is shown that the high frequency tail of the corresponding HN mechanism determines the apparent exponent *n* (*n*<1). Of course, it is not implied by the latter that using the HN function for the description of the polarization process, the UDR behavior can be reproduced.

In the case of Figure 1b, the polarization mechanism follows the Cole-Davidson behavior with HN shape parameters *α*=1 and *β*=0.32. It should be noted that the same mechanism presents a Cole-Davidson behavior at different temperatures.[38] This type of dielectric dispersion is characterized by an asymmetric broadening. The distribution of relaxation times, $G(\tau)$, is highly asymmetric with $G(\tau)=0$ at $\tau > \tau_o$ and $G(\tau) = (\sin\beta\pi/\pi)\tau^\beta/(\tau_o - \tau)^\beta$ at $\tau < \tau_o$, where the characteristic time $\tau_o$ corresponds to the cut-off frequency $f_{HN}$=637 Hz (Figure 1b). For $\tau > \tau_o$ the manifestation of DC conduction takes place (DC conductivity plateau). The high frequencies slope *n* for $\tau \ll \tau_o$ depends on the HN parameter *β*, according to Equation (8), which defines the broadening of the $G(\tau)$ function. A higher value of *β* implies a narrow distribution of relaxation times and consequently a lower value of the power exponent *n* (and vice versa). Although the polarization mechanism shown in Figure 1b exhibits a Cole-Davidson behavior at different temperatures,[38] the results of the present study cannot assure that it is a universal behavior of this slow polarization process in polymeric materials based on the findings of the present study.

When $\sigma_o < \varepsilon_o \omega \varepsilon_d''(\omega)$ at frequencies $\omega \leq \omega_{HN}$, for several orders of frequency magnitude, then Equation (2) via Equation (7) is written as:

$$\sigma'(\omega) \cong \sigma_o + \varepsilon_o \omega_{HN}^{-\alpha} \Delta \varepsilon \beta \sin(\alpha \pi / 2) \omega^{1+\alpha} \qquad (9)$$

in the frequency range $\omega \leq \omega_{HN}$. In this case, the power law dependence of the real part of complex electrical conductivity, as Equation (9) shows, characterized by an apparent exponent *n* which is equal to $1+\alpha$ and it should take values in the range $1<n\leq 2$. From Equation (9) is evident that, when the dielectric dispersions obey Cole-Davidson or Debye behavior ($\alpha=1$), then the high frequency slope could obtain a maximum value *n*=2.

Figure 2a shows the frequency dependence of AC conductivity for the poly(methyl methacrylate) (PMMA) sample at 170 $^o$C. From the low frequency plateau the DC conductivity is found to be $5.8 \times 10^{-10}$ S/m, while the high frequency slope is *n*=1.67 in the frequency range 5-500 kHz. Figure 2a shows also the imaginary part of dielectric constant as a function of frequency for PMMA at the same temperature. The high $\varepsilon''$ values at the low frequency side, with a slope equal to -1, are due to the contribution of the DC conductivity. The strong high frequency dispersion is the dipolar *β*-relation related to the hindered rotation of the ester side groups attached to the main chain.[41] The weaker *α*-relaxation mechanism, which is associated with the glass transition, comes into the frequency window and merges very fast with the particularly strong *β*-relaxation at temperatures higher than 125 $^o$C.[41] Without damaging generality, the effect of the weaker *α*-relaxation has been ignored, because the dielectric losses it describes are relatively small compared to those of the *β*-relaxation mechanism. From Figure 2a it is apparent that at frequencies higher than the crossover frequency of 1.5 kHz, the DC conductivity is lower than the term $\varepsilon_o \omega \varepsilon_d''(\omega)$. Also, the *β*-relaxation exhibits frequency dependence as $\varepsilon'' \propto \omega^{0.67}$ at the low frequency regime of 5-500 kHz. Thus, Equation (9) could be considered as an equation describing the frequency dependence of AC conductivity. Figure 2a includes the best fitting curve according to Equation (9), by keeping constant the values $\sigma_o=5.8 \times 10^{-10}$ S/m and α=0.67. The other parameters were found to be: β=1, Δε=1.34 and $f_{HN}= \omega_{HN}/2\pi=1.31$ MHz. The fitting curve based on Equation (9) describes very well the experimental data, despite the omission of *α*-relaxation contribution. As shown in Figure 2a, the apparent power exponent is *n*=1.67, but this superlinear value could not be related to the high frequency charges motion, because it is connected to the HN shape parameters of a molecular dipolar relaxation. It has been shown that the analysis of frequency dependence of the real part of complex electrical conductivity data can provide an apparent exponent *n*>1, when molecular dipolar relaxation process contributes to the dielectric response in the recorded frequency window while the process has a mean time scale much shorter than the experimental one. In this case the low frequency tail of the corresponding HN mechanism determines the apparent exponent *n* (*n*>1).

Figure 2b shows a simulation curve according to the combination of Equations (2) and (4), where the fitting parameter values of Equation (9) in Figure 2a were used. The slope in $\log \sigma'(\omega) - \log f$ receives know a value of 1.56, in the range 5-500 kHz. At higher frequencies, $f>f_{HN}$, the slope gradually decreases as shown in Figure 2b, while the *n* exponent approaches the value of 0.33 in the 3-5 GHz range. A similar behavior presents the second simulation curve, which is also based on the combination of Equations (2) and (4). In the later simulation the parameter values $\sigma_o$, α, *β* and *Δε* were kept constant, while the value of 13.1

MHz was selected for $f_{HN}$, which is one order of magnitude higher than $f_{HN}$ of the first simulation in Figure 2b. In the later simulation curve the slope receives the higher value of 1.61 in the range of 5-500 kHz, while for the highest frequency range of 3-5 GHz (Figure 2b) the apparent *n* exponent receives the value of 0.35. The value of intermediate slope in the range 5-500 kHz increases as $f_{HN}$ is shifted to higher frequencies. Figure 2b also includes for comparison the fitting curve shown in Figure 2a, which is expressed via the relation $\sigma'(f) = 5.8 \cdot 10^{-10} + 5.16 \cdot 10^{-15} f^{1.67}$, with a slope of value 1.67 in the 5-500 kHz range. A slight deviation in the value of apparent *n* exponent is observed as calculated in the first simulation (Figure 2b) by using the same parameter values. The difference in the values of apparent *n* exponent is due to the fact that, for the first simulation curve, the parameter values calculated from the best fit of Equation (9) were used in Equations (2) and (4). It should be noted that Equation (9) applies to the frequency range $f<<f_{HN}$, according to Equation (7). However, in the frequency range where apparent *n* exponent was calculated, the condition $f<<f_{HN}$ =1.31 MHz is not satisfactory, therefore the combination of Equations (2), (4) and Equation (9) are not identical and consequently the values of the apparent *n* exponent will be different, but without much divergence. Besides commenting on the findings about intermediate slope, the aim of Figure 2b is to show that the combination of Equations (2) and (4) lead to a value of the apparent *n* exponent that approximates the value 1-*αβ* for frequencies $f>>f_{HN}$. When the polarization mechanism is a Debye type (*α=β*=1), then saturation occurs at higher frequency range ($f>>f_{HN}$), in the $\log \sigma' - \log f$ plot.

Figures 1 and the corresponding text describe the study conducted on the case of the DC conduction in disordered systems, such as in polymeric materials, which is characterized by the existence of a slow polarization mechanism due to the short range charges motion. Furthermore, Figures 2 and the corresponding text delineate the examination that was carried out in the case of DC conduction in polymeric materials with an additional dominant molecular dipolar relaxation, which eliminates the effect of slower polarization processes in the formation of high frequency slope of AC conductivity. In what follows, will be examined the case where except the DC conduction, an additional molecular dipolar relaxation appears at higher frequencies, which restricts the effect of slow polarization process in the formation of the apparent power exponent $n$.

The polarization mechanisms due to short range motion of charges are slower processes compared to any molecular dipolar dispersion at the same polymeric material. We now consider a situation where, apart from DC conductivity $\sigma_o$, two dielectric loss mechanisms are also present: the first one at the low frequency regime having a characteristic frequency, $\omega_{HN,l}$, where $\omega_{HN,l}$ is some orders of magnitude lower than the characteristic frequency of the second one which is $\omega_{HN,h}$. The lower frequencies mechanism is assumed to include the entire contribution of the mobile ions effects and applies for $\sigma_o \geq \varepsilon_o \omega \varepsilon_d''(\omega)$ at $\omega \leq \omega_{HN,l}$. The higher frequencies dispersion is assumed as a molecular dipolar mechanism with $\sigma_o < \varepsilon_o \omega \varepsilon_d''(\omega)$ at $\omega \leq \omega_{HN,h}$ for several orders of frequency. Also we consider that, between frequencies $\omega_{HN,l}$ and $\omega_{HN,h}$ a well defined shallow minimum in $\varepsilon''$ is formed. Then, between $\omega_{HN,l}$ and crossover frequency, the losses of the mechanism characterized by $\omega_{HN,l}$ is dominant and the slope in $\log\sigma' - \log\omega$ plot is mainly determined by the contribution of $K_l \omega^{1-\alpha_l \beta_l}$, where the factor $K_l$ equals to the multiplier of quantity $\omega^{1-\alpha\beta}$ in Equation (8). In the range between the crossover frequency and the frequency $\omega_{HN,h}$, the losses of mechanism characterized by $\omega_{HN,h}$ dominate and the slope is determined mainly by the contribution of $K_h \omega^{1+\alpha_h}$, where $K_h$ equals to the multiplier of quantity $\omega^{1+\alpha}$ in Equation (9).

Figure 3a shows the variation of $\varepsilon''$ as a function of frequency for the contributions of DC conductivity and two dielectric loss mechanisms. For the low frequency dispersion and DC conductivity contribution the parameters were kept constant as the values of Figure 1: $\alpha_l = 1$, $\beta_l = 0.32$, $\Delta\varepsilon_l = 4.63$, $f_{HN,l} = 637$ Hz, and A=6260,24 in term $Af^{-1}$. Also for the faster relaxation only the HN shape parameters were changed as shown in Figure 3a, while the other were kept constant: $\Delta\varepsilon_h = 2$, $f_{HN,h} = \omega_{HN}/2\pi = 10$ MHz, while the parameter $\alpha_h$ has been considered to vary by taking values 1, 0.7 and 0.5. Although the case with $\alpha=0.5$ does not corresponds to a well defined shallow minimum in $\varepsilon''$, it is mentioned here just to indicate how an apparent NCL situation can be formed. The curve with $\alpha_h = 0.5$, which represents a broad Cole-Cole dielectric behavior, exhibits almost constant losses for about 1.5 orders of magnitude, between 0.5-10 MHz, as it shown in Figure 3a. From Figure 3b, which gives the frequency dependence of AC conductivity for the corresponding simulation curves at $f < f_{HN,h}$, it becomes evidence that the case with $\alpha_h = 1$ which represents a Cole-Davidson behavior exhibits the largest value of apparent $n$ exponent which is equal to 1.51. The case where $\alpha_h = 0.7$ is the first Cole-Cole mechanism and it is characterized by a high frequency slope equal to 1.18. It is obvious that, the superlinear values could not be related to the high frequency ions motion, because they are connected to the HN shape parameters of a molecular dipolar relaxation. The second case with the broad Cole-Cole behavior ($\alpha_h = 0.5$) characterizes the well known NCL behavior with a high frequency slope equal to 1.02, as expected from the $\log\varepsilon'' - \log\omega$ plot of Figure 3a. Therefore, between two well-separated dielectric mechanisms with clear shallow $\varepsilon''$ minimum, the high frequencies slope can take values higher than 1. It is observed that, with decreasing value of the shape parameter $\alpha$, the superlinear value of the apparent $n$ exponent, which is calculated in the frequency range of the one order of magnitude lower than $f_{HN}$, gradually decreases. The high frequencies slope of the AC conductivity is mainly formed from the contribution of slower components of the molecular dipolar relaxation at this frequency range. A narrow distribution of relaxation times of slower component of the dipolar relaxation and consequently a higher value of the shape parameter *α*, leads to a higher value of the apparent *n* exponent (and vice versa). Also, as it has been shown above, the proper coupling of the dielectric mechanisms HN parameter values can lead to an apparent NCL in a relatively extended frequencies range, giving rise a high frequency slope very close to 1.

## 4. Conclusions

Summarizing, the AC conductivity response is defined by, and depends on, the dielectric loss mechanisms. Thus, the dynamic characteristics of dielectric loss mechanisms shape entirely the behavior of $\sigma'$ in the whole frequency spectrum. Depending on the correlation of the DC conductivity as well as the dynamic characteristics of dielectric loss mechanisms, the apparent *n* exponent receives values less than 1. In some cases, the apparent *n* exponent could reach values in the range $1 \leq n \leq 2$ for a relatively narrow frequencies window as compared to whole spectrum in the classical dielectric region. In the case of the existence of DC conductivity and various dielectric dispersions, the high frequency slope of the $\log\sigma'(\omega) - \log\omega$ plots of the whole frequency spectrum approaches values lower than 1 and equal to $1 - \alpha\beta$, where *α*, *β* are the HN shape parameters of the highest frequencies mechanism.

The DC conduction in disordered materials, such as polymers, seems to be connected to the existence of a polarization process, which is related to a short range motion of mobile charges. These mobile charges contribute to the DC conductivity at low frequency regime. The importance of the *n* slope in various interpretations, applications and models, lies in its relation to this polarization mechanism which includes an entire contribution due to the mobile charges effects. Therefore, this mechanism is the slowest dielectric process. In the case of the appearance of DC conduction and the previous polarization mechanism in $\log\varepsilon'' - \log\omega$ plot, the apparent power exponent $n$ receives a value of $1-\alpha\beta$ and is directly related to the characteristics of charges motion at shorter time scales. When an additional molecular dipolar relaxation contributes at higher frequencies in $\log\varepsilon'' - \log\omega$ plot, then the apparent power exponent could take values in the range $1 \leq n \leq 2$ for a relatively narrow frequency range.

The existence of a well defined minimum between DC conductivity contribution and a dielectric dispersion or between two well separated dielectric loss mechanisms, in $\log\varepsilon'' - \log\omega$ plots, give rise to the appearance of apparent $n$ values in the range $1 < n \leq 2$, above the crossover frequency. As the molecular dipolar relaxations are the faster processes in polymers, the previous superlinear apparent values of $n$ should not be taken into consideration, because these values could not be related to the characteristics of mobile charges at high frequency region. In these cases, the apparent power exponent $n$ is simply related to the HN shape parameter $\alpha$ of the higher frequencies molecular dipolar relaxation. The findings of the present study could also be extended to other disordered systems as well in the case where their dielectric loss mechanisms can be described by the HN equation.

**Acknowledgments**

Many thanks to my old colleagues: to Dr Apostolos Kyritsis for the helpful discussion, and to Dr Emmanuel Logakis for supplying to me the PMMA experimental dielectric data.

**Figure captions**

**Figure 1.** Dielectric response of a polyurethane sample at 90 °C. (a) Frequency dependence of AC conductivity: experimental data (green line) and simulation curve (red line) calculated following Equation (8). Data were taken from Figure 5 of Reference [38]. (b) The imaginary part $\varepsilon''$ as a function of frequency (green line) with the contribution of the polarization mechanism and DC conductivity (red lines).

**Figure 2.** (a) Dielectric response for experimental data of poly(methyl methacrylate) (PMMA) sample at 170 °C. The data were given to the author by Dr. E. Logakis. Blue circles show the frequency dependence of AC conductivity while blue line is the best fitting according to Equation (9). (b) Simulation curves of AC conductivity based on the combination of Equations (2) and (4) and the previous fitting curve according to Equation (9).

**Figure 3.** Simulation curves including the contributions of DC conductivity and two dielectric loss mechanism according to Equations (2) and (4) (details in the text). (a) Frequency dependence of the imaginary part $\varepsilon''$. (b) Frequency dependence of AC conductivity at f<$f_{HN,h}$. The inset shows this more clearly at the high frequencies region.

**Figure 1**

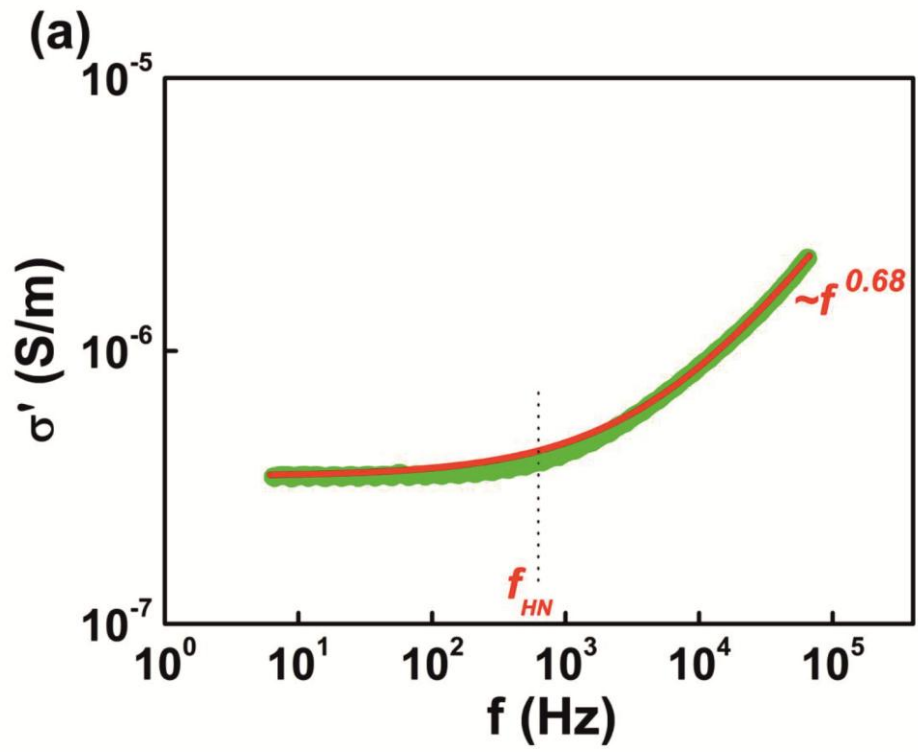

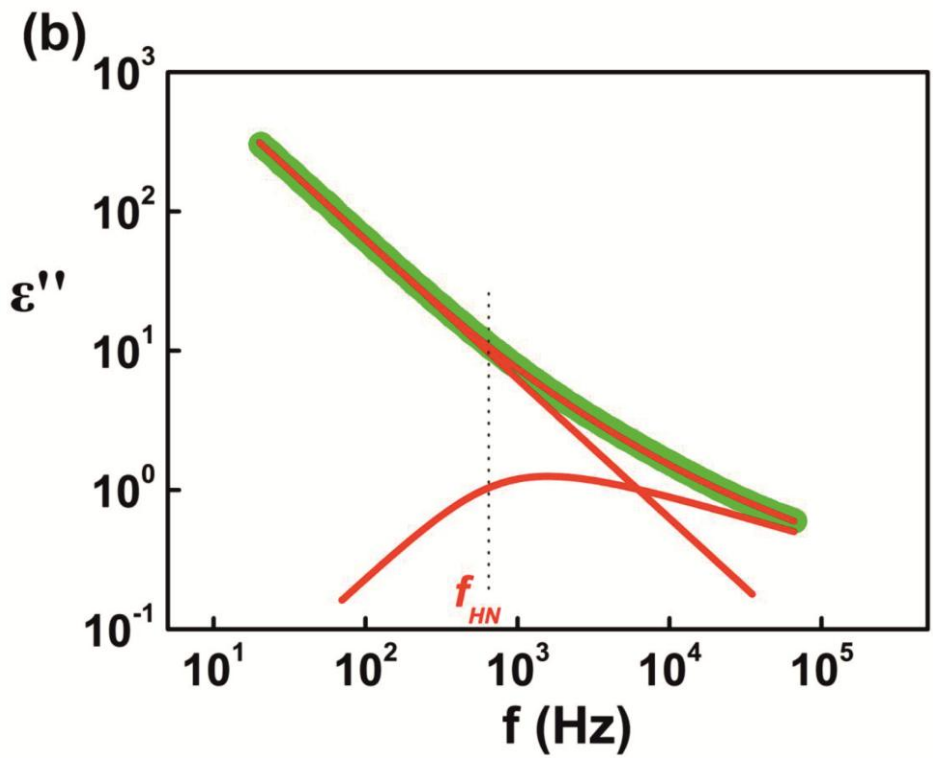

**Figure 2**

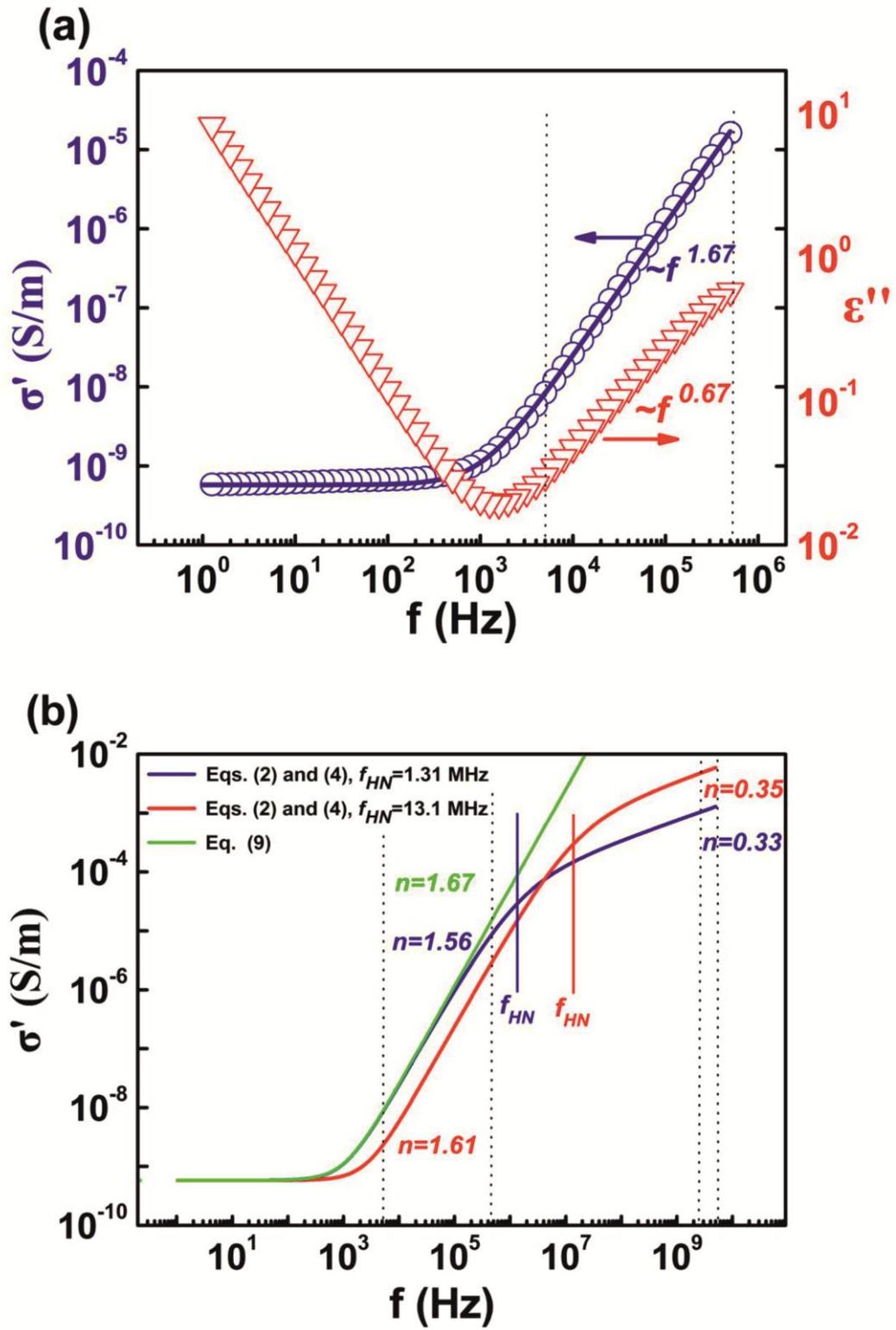

**Figure 3**

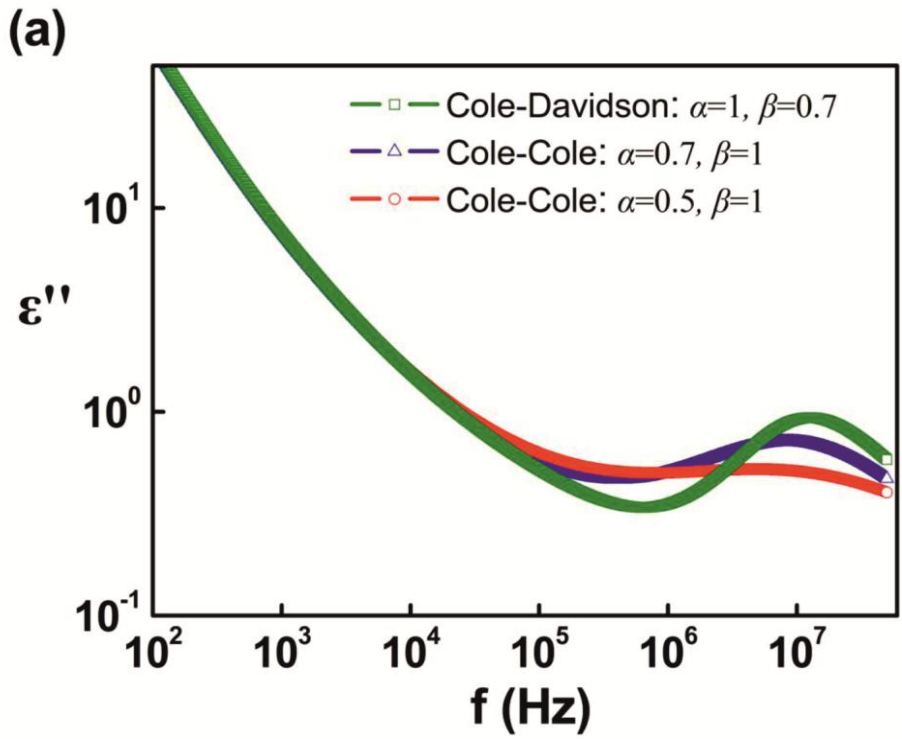

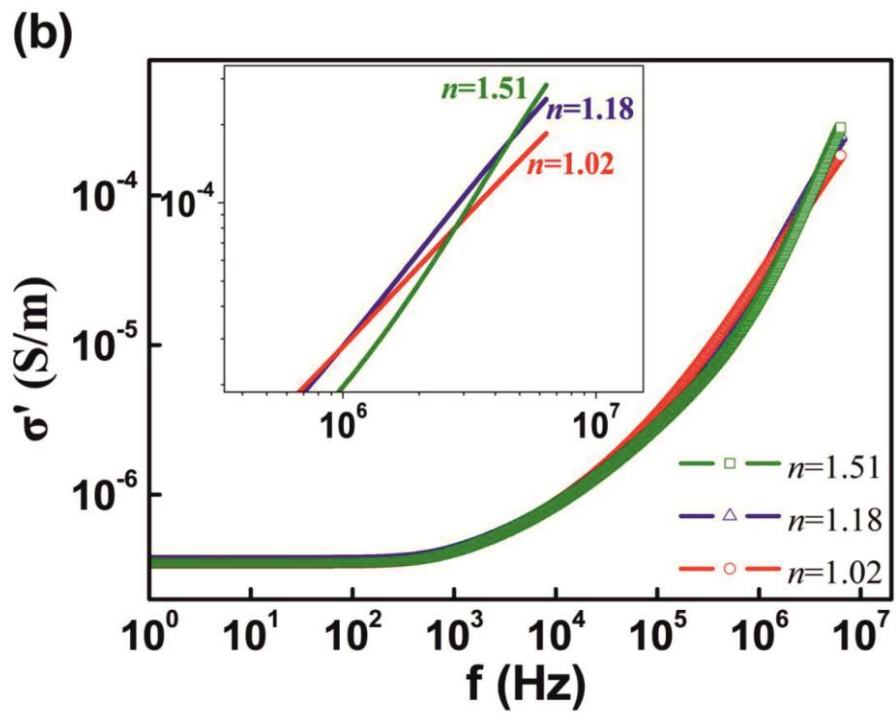